\documentclass[review,5p,twocolumn]{elsarticle}
\usepackage{lineno,hyperref}
\usepackage{amsmath,amssymb,latexsym}
\usepackage{wrapfig}
\usepackage{subfig}
\usepackage{times}
\usepackage{epsfig}
\usepackage{graphicx}% Include figure files
\usepackage{color}
\usepackage{dcolumn}% Align table columns on decimal point
\usepackage{bm}% bold math
\usepackage{setspace}
\usepackage{enumitem}
\usepackage{caption}
\usepackage{siunitx}

\newcommand \nline {\nonumber \\}

\newcommand \pxpy[2] {\frac{\partial #1}{\partial #2}}

\newcommand \E[1] {Eq.~(\ref{#1})}
\newcommand \Es[1] {Eqs.~(\ref{#1})}

\newcommand \be {\begin{equation}}
\newcommand \ee {\end{equation}}
\newcommand \ben {\begin{eqnarray}}
\newcommand \een {\end{eqnarray}}

\newcommand{\beq}{\begin{equation}}
\newcommand{\eeq}{\end{equation}}
\newcommand{\beqa}{\begin{eqnarray}}
\newcommand{\eeqa}{\end{eqnarray}}

\newif\ifbw
\bwfalse

\usepackage{amsmath}
\usepackage[toc,page]{appendix}
\usepackage{float}
\usepackage{subfig}
\usepackage{graphicx}
\usepackage{verbatim}
\usepackage{cancel}
\usepackage{color}
\usepackage[]{algorithm2e}
\usepackage{listings}

\definecolor{dkgreen}{rgb}{0,0.6,0}
\definecolor{gray}{rgb}{0.5,0.5,0.5}
\definecolor{mauve}{rgb}{0.58,0,0.82}

\begin{document}
%%%%%%%%%%%%%%%%%%%
%\begin{frontmatter}
\title{Quantitative phase field modeling of solute trapping and continuous growth kinetics in rapid solidification}
\author[1,2]{Tatu Pinomaa}
\author[2]{Nikolas Provatas}
\address[1]{VTT Technical Research Centre of Finland Ltd, Espoo, Finland}
\address[2]{Department of Physics and Centre for the Physics of Materials, McGill University, Montreal, Canada}

\begin{abstract}
%Solute trapping is an important phenomenon in rapid solidification of alloys. The continuous growth model (CGM) is a popular description for sharp interface theories of this solidification regime. 
Solute trapping is an important phenomenon in rapid solidification of alloys, for which the continuous growth model (CGM) is a popular sharp interface theory. 
Using matched asymptotic analysis, we show how to quantitatively map the sharp interface behavior of a binary alloy phase field model  onto the CGM kinetics of Aziz et al. \cite{Aziz94}, with a controllable partition coefficient $k(V)$.
% of a binary alloy phase field model, we show how to control solute partitioning
%We show how to map the behavior of the phase field model onto a given 
%that a phase field model can achieve a specific non-equilibrium partition coefficient $k(V)$ together with kinetic undercooling according to the continuous growth model. 
%The phase field model can be set up to follow continuous growth model either with full or vanishing solute drag. 
%By extrapolating the phase field model concentration to the interface on solid and liquid sides, %transient planar 
%For different values of diffuse interface widths,
We demonstrate the parameterizations that allow the phase field model to map onto the corresponding CGM or classical sharp interface models. 
We also demonstrate that the mapping is convergent for different interface widths.
Finally we present the effect that solute trapping can have on cellular growth in a directional solidification simulation. 
The treatment presented for solute trapping can be easily implemented in different phase field models, and is expected to be an important feature in future studies of quantitative phase field modeling in rapid solidification regimes, such as those relevant to additive manufacturing.
\end{abstract}
\date{\today}

\maketitle

\section{Introduction}
Rapid solidification of metallic alloys is a common feature in advanced industrial manufacturing processes such as additive manufacturing, laser welding, and thermal spray coatings. The rapid solidification is often accompanied with incomplete solute partitioning at the solid-liquid interface, which is called \textit{solute trapping}. This affects the solidification microstructure by influencing the growth morphology, length scale, microsegregation and the resulting precipitation of secondary phases. These microstructural features determine, to a large extent, the properties and performance of the material. Moreover, these features can be related back to the controllable process details through computer modeling.

Classic sharp interface models (hereafter {\it SIM}) can well be used to describe traditional casting processes. operate at low to moderate cooling rates, which are well described by the classic sharp interface model (hereafter {\it SIM}), typically with a vanishing kinetic coefficient ($\beta=0$). 
%Traditional casting processes operate at low to moderate cooling rates, which are well described by the classic sharp interface model (hereafter {\it SIM}), typically with a vanishing kinetic coefficient ($\beta=0$). 
The classical SIM assumes zero interface width, and that the interface is near equilibrium during solidification. This practically means that the solid-liquid interface is much smaller than the capillary length, which is the smallest characteristic length in the solidification problem. During rapid solidification, in contrast, the equilibrium conditions that prevail in the classic SIM break down, and the atomic attachment kinetics and other non-equilibrium effects that emerge in the physically non-zero solid-liquid interface. These include a velocity-dependent solute partition coefficient $k(V)$ and a velocity-dependent interface undercooling, or concentration. These effects become dominant at rapid solidification  rates and can strongly affect the microstructure kinetics, morphology and phase formation. 

A convenient method for modeling microstructure problems in solidification and solid-state transformation is phase field method.
%Phase field modeling is increasingly being used to solve a multitude of microstructure problems in solidification and solid-state transformations.
This is due to its fundamental origins, connections with non-equilibrium thermodynamics, and numerical efficiency compared to interface tracking approaches.  Phase field modeling has been used in the study of solidification in a range of materials, from ideal dilute binary alloys \cite{Karma01,Echebarria04} to more complex binary alloys \cite{Plapp11} and multi-component or multi-phase alloys \cite{Eik06,Nestler12,Greenwood18}.

For a special class of models called multi-phase field  models, Steinbach et al. presented a finite dissipation model \cite{steinbach2012} for simulating solute trapping in rapid solidification. A specific solute partitioning $k(V)$ is achieved by coupling a kinetic equation between the phase concentration, and by adjusting a so-called {\em rate constant} together with the numerical interface width to control $k(V)$ through the interface dissipation term. Their results for $k(V)$ are are consistent with a different approach of Danilov and co-workers \cite{danilov2006} in a similar range of interface velocities, and both are consistent with experiments. These studies did not report phase field model predictions of kinetic interface concentration or undercooling, and how these compare to predictions of a non-equilibrium solidification model, such as the  continuous growth mode (CGM) of Aziz and Boettinger \cite{Aziz94}.

This paper examines the continuous growth limit of another class of phase field models based on order parameter fields \cite{Echebarria04}. In the limit of low undercooling (or low supersaturation), a  robust set of results derived from a matched asymptotic  boundary layer analysis of this model, for an ideal binary alloy \cite{Echebarria04}, can be used to  map the model's behaviour {\em quantitatively} onto the classical sharp interface model; these results can also be essentially used to recover the classical sharp interface limit of most of the above-cited phase field models \cite{Plapp11,Nestler12,Greenwood18}.  

%Rapid solidification is becoming the norm in advanced industrial manufacturing processes such as additive manufacturing, laser welding, and thermal spray coatings. Yet for this solidification regime, 

Previous order parameter-based phase field model studying rapid solidification have used the aforementioned classical sharp interface limit (i.e. $k(V) = k_e$ in most studies \cite{kundin2015,wu2018,sahoo2016}. For the same classical SIM parametrization, Ghosh et al. \cite{ghosh2018} includes solute trapping by using a combination of large $W$ and $V$, such that incomplete anti-trapping leads to some emergent $k(V) > k_e$, which  depends on the chosen interface width $W$, and is also different in $1D$, $2D$, and $3D$ simulations.  Currently no quantitative phase field model parameterization exists which consistently maps a phase field model onto the appropriate non-equilibrium sharp interface limit described by a specific $k(V)$ and interface undercooling/concentration.

There is presently no generally accepted SIM to describe the rapid solidification regime. Several sharp interface models for this regime have been proposed. The two most popular paradigms are the {\it continuous growth model} (CGM) of Aziz and co-workers \cite{Aziz88,Aziz94} and that of Sobolev and co-workers \cite{sob95,sob97}. The former assumes standard  diffusion accompanied by attachment-limited kinetics at the interface, while the latter further incorporates two-time-scale dynamics to  describe both inertial and diffusive dynamics of solute atoms near and through a rapidly advancing interface. The two approaches give similar results at low velocities (although still large enough to be in the rapid solidification regime). In this work, we will focus on the former, however, we expect our results to be straightforwardly generalizable to the latter. 

Ahmad et. al \cite{Ahmad98}, Wheeler et. al \cite{Wheeler93} and Boettinger et. al \cite{Boettinger94} showed that a phase field model of alloy solidification, governed by first order diffusion kinetics, captured most of the salient features of the continuous growth model of Aziz and co-workers. However, these works also found that the fundamental parameters of any effective CGM projected out of a phase field model (e.g.  the segregation coefficient $k(V)$ and kinetic undercooling) are  sensitive to the phenomenological interpolation functions that are designed into the original phase field equations.  Moreover, the connection between the two models is non-trivial, making the description of the physics of rapid solidification difficult to do {\it quantitatively}.

This paper will show how to systematically map a binary alloy phase field model containing an anti-trapping flux onto the continuous growth model (CGM) model described by a specific form of the solute trapping coefficient $k(V)$ and  kinetic interface undercooling.  This is presented  here for the classic case of an ideal binary alloy. However the results of the general matched asymptotic analysis, presented in the supplementary material,  are easily generalized by working out new coefficients for non-dilute and multicomponent alloys.  

The paper begins by summarizing the continuous growth model of rapid solidification in the limit of a sharp solid-liquid interface. This is followed by a summary of the standard ideal dilute binary alloy phase field model, which uses a non-variational formulation with a so-called anti-trapping current \cite{Echebarria04}. The results of  a matched asymptotic analysis of this model, extracted from the supplementary materials, are used to demonstrate how the aforementioned phase field model can be parameterized to simulate a specific form of $k(V)$ and the kinetic interface concentration described by the CGM. Both the CGM limits corresponding to full solute drag and zero solute drag are considered.  For comparison, we show the equilibrium partitioning $k(V) = k_e$ with kinetic coefficient $\beta$ set to either zero or to an experimentally relevant value.
\section{Methods} 
This section briefly reviews the  continuous growth model in the sharp interface limit, and the ideal dilute binary alloy phase field model used in this work, and its extension to the CGM regime. 
\subsection{Review of continuous growth model}
\label{section:CGM}
In continuous growth model for dilute binary alloys, the non-equilibrium partition coefficient has the  form \cite{Aziz94} 
\begin{align}
k^{CGM}(V) &= \left( k_e + \frac{V}{V_D^{CGM}} \right)/\left( 1 + \frac{V}{V_D^{CGM}} \right),
\label{eq:k_V_CGM}
\end{align}
where $V$ is the interface velocity,  $k_e$ is the equilibrium partition coefficient, $V_D^{CGM}$ is the so-called diffusive velocity which is typically fit to velocity - partition coefficient experiments. 
%Note that the non-equilbrium partition coefficient for the phase field model, presented in Section \ref{section:PF_soluteTrapping} has a different form than $k^{CGM}(V)$ above.

The continuous growth model also predicts a kinetic undercooling that has a velocity-dependent liquidus slope \cite{Aziz94}. Assuming an externally imposed temperature at the interface, $T$, the kinetic undercooling expression can be inverted to give the liquid-side concentration as %, $ {c_{L,i}}$,  according to 
%%%%%%%%
\begin{align}
 \frac{c_{L}}{c_l^o} & = \frac{1}{f\left(k(V)\right)} \Bigg( 1 + \frac{ T_l - T}{|m_l^e|c_l^o}  -\left(1-k_e\right) d_o \kappa
\nline & 
 - \left( 1-k_e \right) \beta V \Bigg),
 \label{eq:c_L_CGM}
 \end{align}
 %%%%%%%%%
where $c_l^o$ is the average solute concentration in the alloy, $T_l$ is the liquidus temperature, $d_o$ is the solutal capillary length, $\kappa$ is the local interface curvature, $\beta$ is the kinetic coefficient, and $f\left(k(V)\right)$ is the velocity-dependent correction to the liquidus slope, given by 
%%%%%%%
\begin{align}
f\left(  k(V) \right) &= \frac{1}{1-k_e} \Bigg( \left[ k(V)+ \mathcal{D} (1-k(V))  \right] \log \left( \frac{k(V)}{k_e} \right)
 \nline & 
 + 1 - k(V) \Bigg) ,
\label{eq:f_k_V_function}
\end{align}
%%%%%%
where $\mathcal{D}$ is a parameter that can be tuned to represent complete solute drag ($\mathcal{D}=1)$ or no solute drag ($\mathcal{D} = 0$) \cite{Aziz94}. 
%Intuitively, solute drag dissipates part of the energy associated with solidification driving force,
For sufficiently small interface velocities, solute partitioning can be assumed to be at equilibrium, i.e.  $k(V) \rightarrow k_e$. In this limit $f(k(V)) \approx 1$ in \E{eq:f_k_V_function}, and the liquid-side concentration in \E{eq:c_L_CGM} becomes the classic Gibb Thomson condition for binary alloys.

Note that  \E{eq:f_k_V_function} generally holds for any non-equilibrium partition coefficient $k(V)$, not just the form given by \E{eq:k_V_CGM}. It is thus expected that simulating the CGM limit in phase field simulations should also allow for independent control of the partition coefficient and kinetic undercooling.

%
% T_i &= T_m -  f(k) |m_l^e| c^l - \frac{1}{\mu_k} V,  This can be rewritten in terms of concentration as  where the solutal kinetic coefficient $\beta := 1/\left( (1-k_e) |m_l^e| c_l^o \right)$

\subsection{Phase field model of an ideal binary alloy}
\label{section:PF_standard}
Phase field modeling of solidification of a dilute binary alloy is described by an order parameter $\phi$ (using here the limits  $-1 \leq \phi \leq 1$)  and concentration field $c$, whose dynamics are governed by 
%%%%
\begin{align}
\tau \pxpy{\phi}{t}
& =  \nabla \cdot \left[ W^2 \nabla \phi 
+ W |\nabla \phi |^2 \left( \sum_{k}^{x,y,z}
  \pxpy{W}{ (\partial \phi/\partial k )}\hat{e}_k
  \right) \right]
\nline
 +&  \phi - \phi^3  - \frac{\lambda}{1-k_e} \left( e^u - 1 - \frac{ T_l - T}{|m_l^e| c_l^o} \right) (1-\phi^2)^2,
\label{eq:PF_phi_evolution}
\\
\pxpy{c}{t} = & \nabla \cdot \left[ D_L \,c \,q(\phi)  \nabla u 
 + a_t W_0 (1-k_e) e^u \pxpy{\phi}{t} \frac{ \nabla \phi }{| \nabla \phi |} \right],
\label{eq:PF_c_evolution}
\\
e^u &= \frac{c}{c_{eq}}, \, \, \,
c_{eq} = \frac{ 1+k_e -(1-k_e)h(\phi) }{2},
\nline
h(\phi) &= \phi, \, \, \,
q(\phi) = \left( \frac{ 1 - \phi}{2} + \frac{1 + \phi}{2} \frac{ D_S}{D_L} \right)/c_{eq},
\label{eq:h_q_interpolation}
\end{align}
%%%%%
where  $\tau=\tau(\mathbf{n})$ is the anisotropic interface attachment time scale, $W=W(\mathbf{n})$ is the anisotropic interface width and $W_0$ is its magnitude, $\lambda$ is the coupling constant, $m_l^e$ is the equilibrium  liquidus slope, $D_{L/S}$ is the liquid/solid diffusion coefficient, and $a_t$ is the antitrapping coefficient.  

\subsection{Classic sharp interface limit of  phase field model}
\label{section:PF_standard}
The sharp interface limit of a phase field model is achieved by matching the perturbed solutions of the phase field equations in the outer region (i.e. beyond the length scale of the diffuse phase field interface) with the asymptotic form of the solutions from the inner region (i.e. on the length scale of the interface). Projecting the matched outer solutions onto the effective interface defined by the $\phi$ field (e.g. where $\phi=0$) yields the boundary conditions of the effective sharp interface model obeyed by the phase field equations, and the parameter relations defining the effective capillary length and kinetic coefficient. The process of projecting the outer solution of the concentration field into the effective sharp interface defined by the midway point of the order parameter field is illustrated in Fig.~(\ref{fig:profileExample}).      

This classic (low undercooling) sharp interface limit of the above phase field model is done  by using the well-established parameter relationships derived in Refs.~\cite{Karma01,Echebarria04}. Namely, the parameters $W$, $\tau$ and $\lambda$ in in Eqs. \ref{eq:PF_phi_evolution} and \ref{eq:PF_c_evolution} are related to the  solutal capillary length $d_o$ and kinetic coefficient $\beta$ according to 
%%%%%
\begin{align}
d_o(\mathbf{n}) &= a_1 \frac{ W(\mathbf{n}) }{\lambda} 
\label{eq:d_o_thinInterfaceRelation}
\\
\beta(\mathbf{n}) &= a_1 \frac{ \tau( \mathbf{n}) }{\lambda \, W(\mathbf{n}) } - a_1 a_2 \frac{ W(\mathbf{n}) }{ D_l},
\label{eq:beta_thinInterfaceRelation}
\end{align}
%%%%%
where $\mathbf{n}:= \nabla \phi/|\nabla \phi|$ is the interface normal, and  $a_1$, $a_2$, and $a_t$ are asymptotic analysis constants that depend on the chosen interpolation functions. For $h(\phi)$ and $q(\phi)$ given by  \E{eq:h_q_interpolation}, they are given by 
%%%
\begin{align}
a_1 &= 0.8839 \label{eq:a_1}
\\
a_2 &\approx 0.6867 \label{eq:a_2_equilbirium}
\\
a_t &= \frac{1}{2\sqrt{2}} \label{eq:a_t_equilibrium}
\end{align}
%%%%%

The capillary length $d_o(\mathbf{n})$ and kinetic coefficient $\beta(\mathbf{n})$ are typically anisotropic in 2D and 3D. For example, for  cubic crystal lattices with weak anisotropy, this anisotropy is expressed as 
%%%%%%%%%
\begin{align}
d_o(\mathbf{n})/d_o^{mag} &= 1 - 3\epsilon_c + 4\epsilon_c \left( n_x^4 + n_y^4 + n_z^4 \right), 
\label{eq:d_o_anisotropy}
\\
\beta (\mathbf{n})/\beta_0 &= 1 + 3\epsilon_k - 4\epsilon_k \left( n_x^4 + n_y^4 + n_z^4 \right),
\label{eq:beta_anisotropy}
\end{align}
%%%%%%
where $d_o^{mag}$ is the magnitude of the anisotropic capillary length $d_o(\mathbf{n})$, and $\epsilon_c$ is the capillary anisotropy strength. Analogously, $\beta_0$ is the magnitude of the anisotropic kinetic coefficient $\beta(\mathbf{n})$, and $\epsilon_k$ is the kinetic anisotropy strength. 

\subsection{CGM sharp interface limit of phase field model}
\label{section:PF_soluteTrapping}
In this section we will show how the above standard binary phase field model can be modified to model the kinetics of the continuous growth model, described in particular by a particular partition coefficient $k(V)$ and kinetic undercooling given by the CGM model. This will be done by modifying the original form of the antitrapping current $a_t$, which leads to a correction to the asymptotic constant $a_2$.

To show how to achieve controlled solute trapping in the phase field equations in Eqs. (\ref{eq:PF_phi_evolution}) and (\ref{eq:PF_c_evolution}), the antitrapping coefficient $a_t$ in \E{eq:a_t_equilibrium} is modified as follows: 
%%%%
\begin{align}
a_t \rightarrow a_t' &= \frac{1}{2 \sqrt{2}} \left( 1 - A \left( 1-\phi^2 \right) \right),
\label{eq:a_t_soluteTrapping}
\end{align}
%%%
where $A$ is \textit{trapping parameter}, introduced to control the amount of solute trapping. 
As shown in Supplementary material, the modified antitrapping coefficient $a_t'$ in \E{eq:a_t_soluteTrapping}) leads to a modification to the asymptotic analysis constant $a_2$ used to set $\beta$ in \E{eq:beta_thinInterfaceRelation}, given by 
%%%%
\begin{align}
a_2 \rightarrow a_2^{\pm} &= \frac{ J}{\sigma_\phi}\left( \bar{K} + \bar{F}^{\pm}  \right) \label{eq:a_2_nonEq} ,
\end{align}
%%%%
where $a_2^+$ corresponds to zero solute drag, $a_2^-$ corresponds to full solute drag, and the constants in \E{eq:a_2_nonEq} are given by 
%%%%
\begin{align}
\bar{K} & \approx 0.0638 - 0.0505 A,
\nline
J &= \frac{16}{15},
\nline
\sigma_\phi &= \frac{ 2 \sqrt{2} }{3},
\nline
\bar{F}^+ &= \frac{ \sqrt{2} \ln 2 }{2} - \frac{\sqrt{2} }{4} A
\nline
\bar{F}^- &= \frac{ \sqrt{2} \ln 2 }{2} +  3 \frac{\sqrt{2} }{4} A.  \label{a2_constants}
\end{align}
%and where $\alpha$ is used to control the form of solute drag, where the full solute drag limit is obtained with $\alpha = 1$, and the zero solute drag limit at $\alpha = 0$. 

For $A=0$ the modified antitrapping coefficient $a_t'$ reverts back to $a_t$ in \E{eq:a_t_equilibrium} and $a_2'$ reverts back to $a_2$ in \E{eq:a_2_equilbirium}, reducing the phase field model back to the equilibrium model with $k(V) = k_e$. The asymptotic analysis with this new (or any other) form of anti-trapping flux does not change the value of $a_1$ in \E{eq:a_1} and thus the phase field parameterization  of the capillary  length in \E{eq:d_o_thinInterfaceRelation} remains same. 

It is noteworthy that the form of $a_t^\prime$ is a convenient choice that makes the integrals arising from the asymptotic analysis easily  tractable. Other similar forms are possible, each leading to a different specific value of the constants appearing in \E{a2_constants}.

As shown in supplementary material, when the constants $\bar{F}^+\ne \bar{F}^-$ in \Es{a2_constants}, there is a chemical potential jump across the effective sharp interface. It is well documented that this leads to solute trapping as the interface is no longer able to maintain local equilibrium \cite{Echebarria04,ProvatasElder10}. 
To second order in the perturbation theory used to analyze the phase field equations, 
the solute partition coefficient is given implicitly by a transcendental relationship between interface velocity and non-equilibrium partition coefficient:
%%%%%
\begin{align}
k^{PF}(V) &= k_e \exp \left( \sqrt{2} \left(1 - k^{PF}(V)\right)\, V/V_D^{PF}  \right),
\label{eq:k_V_PF}
\end{align}
%%%%%
where
%%%%
\begin{align}
V_D^{PF} &=  \frac{D_L}{A W_0},
\label{eq:V_D_PF}
\end{align}
%%%%
is a characteristic solute trapping velocity, $W_0$ is the magnitude of anisotropic interface width $W(\mathbf{n})$, and $A$ is the trapping parameter for $a_t'$ introduced in \E{eq:a_t_soluteTrapping}. Equation (\ref{eq:k_V_PF}) can be solved numerically, and $V_D^{PF}$ can be chosen to represent a specific amount of solute trapping based on experimental $k(V)$ data. Once an appropriate value for  $V_D^{PF}$ is chosen, the trapping parameter $A$ in \E{eq:a_t_soluteTrapping} is determined through \E{eq:V_D_PF}.
%%%

In addition to a relation for $k(V)$, the asymptotic analysis of \E{eq:PF_phi_evolution} and \E{eq:PF_c_evolution} also predict an equation for the kinetic undercooling of the solid-liquid interface. Specifically, one obtains the following relationships on either the liquid($\ell$) or solid($\alpha$) sides of the effective sharp interface defined by the order parameter,   
%%%%%%%
\begin{align}
\bar{f}_\alpha \left( c^{\ell}  \right)&\, -\, \bar{f}_\ell \left(c^\alpha \right) \,+\, \left( c^{\alpha} - c^{\ell} \right) \, \frac{ \partial \bar{f}_\vartheta( c^{\vartheta} ) }{ \partial c}= -\frac{V}{ v_c },
\label{undercooling}
\\
\text{where } & \vartheta = \ell \, \text{ gives zero solute drag,} \nonumber
\nline
&\vartheta = \alpha \, \text{ gives complete solute drag,} \nonumber
\end{align}
%%%%%%
and $\bar{f}_{\alpha}$ ($\bar{f}_\ell$) is the free energy density of the solid (liquid). The inverse critical velocity ${1}/{v_c}=(1-k_e)^2\,c_o^l \, \beta$, where $\beta$ is given by the following modified sharp interface relation 
%%%%%%%
\begin{align}
\beta(\mathbf{n}) &= a_1 \frac{ \tau( \mathbf{n}) }{\lambda \, W(\mathbf{n}) } - a_1 a_2^{\pm} \frac{ W(\mathbf{n}) }{ D_l},
\label{first_beta2}
\end{align}
%%%%%%
Evaluating Equation (\ref{undercooling})  on the solid side of the interface ($\vartheta = \alpha$), with the phase field parameters set to $a_2^-$  in \E{first_beta2} leads to the CGM model of \E{eq:c_L_CGM} with full solute drag ($\mathcal{D}=1$) \cite{Aziz94}; correspondingly, evaluating \E{undercooling}  on the liquid side and using $a_2^+$ in \E{first_beta2} to set the kinetic time scale of the phase field equations leads to the CGM model of \E{eq:c_L_CGM} with zero drag ($\mathcal{D} = 0$). 

\subsection{Estimating the liquid- and solid-side concentrations from phase field simulations}
\label{section:solidLiquidProjection}
\begin{figure}
\includegraphics[width=0.4\textwidth]{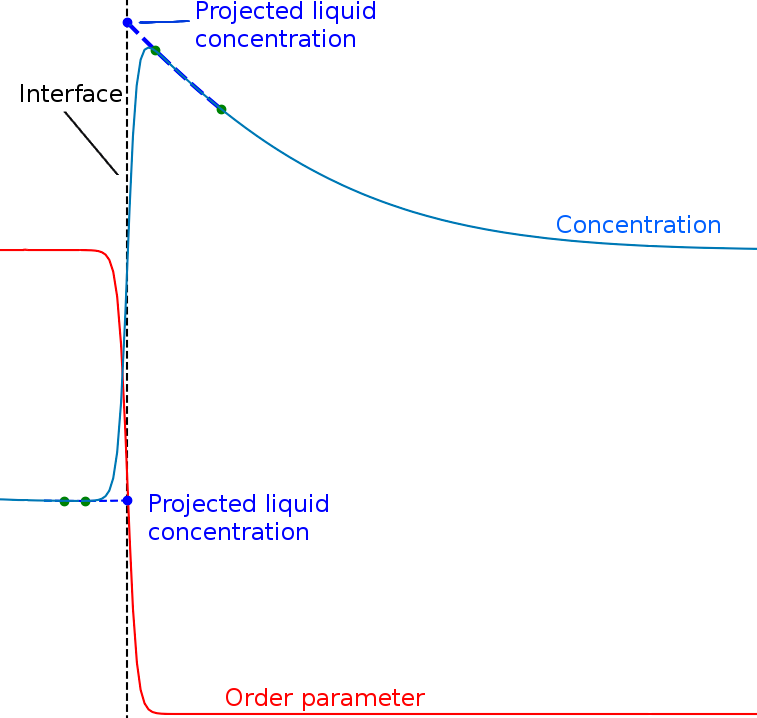}
\caption{Projection of the smoothly varying concentration field (blue) of the phase field model onto the an effective sharp interface interface (vertical dotted line). The projected interfacial concentrations are shown as blue dots.}
\label{fig:profileExample}
\end{figure}
To compare the implemented phase field model to continuous growth model for sharp interfaces, the interfacial solid- and liquid-side concentrations in the phase field model need to be estimated appropriately at the effective interface, defined here by where $\phi=0$.
Our sharp interface estimation of concentration is depicted in Fig.~\ref{fig:profileExample}, where order parameter is the red solid line, and concentration is the blue solid line. For both solid and liquid sides, we fit a second order polynomial to the concentration profile sufficiently far away from the interface (between green dots), where the phase field model's concentration corresponds to the emulated sharp interface model's concentration --- closer to the interface the phase field model's concentration varies smoothly at the interface, whereas the sharp interface model would give out a discontinuous jump at the interface. The fitted polynomial (dashed line) is then extrapolated to the interface to give the interfacial solid- and liquid-side concentrations (blue dots). 

For liquid-side concentration the above approach worked well. However, on the solid-side concentration the above procedure was occasionally corrected manually when the second order polynomial fitting failed. The interface concentration estimation is sensitive to the chosen details of polynomial fitting. %, and leads approximately to a 5\% relative error to the estimated interface concentration values. 
This gave the solid-side estimation of concentration the biggest error, at approximately 5\% relative error --- this estimation error, however did not have a large influence on the evaluation of the partition coefficient. 

We also tested a simpler approach that considered the liquid-side concentration as the phase field profile maximum. With this approach the estimated liquid-side concentration was systematically underestimated compared to the extrapolation approach depicted in Fig~\ref{fig:profileExample}. However, this only slightly affected the error on the results reported below, not the general agreement between phase field simulations and the CGM kinetics.
\section{Results}
\subsection{Determining the solid- and liquid-side concentrations}
\begin{figure}
\includegraphics[width=0.53\textwidth, trim={ 30 0 0 0},clip]{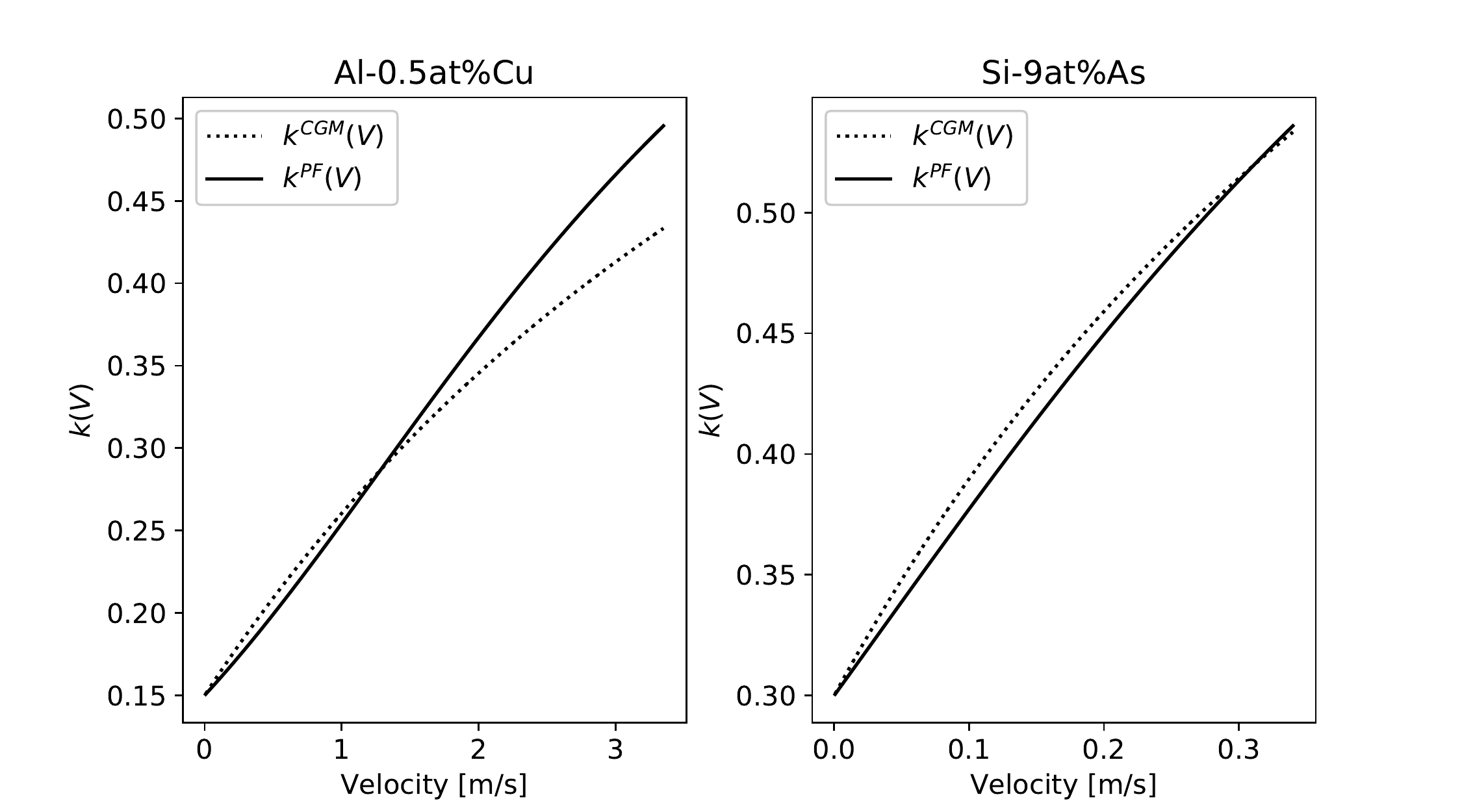}
\caption{Determining the phase field model partition coefficient $k^{PF}(V)$ (solid lines) for Al-Cu alloy (left) and Si-As alloy (right). We adjusted the associated diffusion velocity $V_D^{PF}$ in \E{eq:k_V_PF} to match an experimentally calibrated $k^{CGM}(V)$ from \E{eq:k_V_CGM} (dashed lines). For Al-Cu alloy $V_D^{PF} = 2.0 $ m/s, and for Si-As alloy $V_D^{PF} = 0.385 $ m/s.}
\label{fig:k_V_fitting}
\end{figure}

To determine an appropriate amount of solute trapping in the phase field model,  the characteristic solute trapping speed of the model, $V_D^{PF}$, was adjusted to match the $k(V)$ according to \E{eq:k_V_PF} to an experimentally fitted partition function $k^{CGM}(V)$ as closely as possible at low interface velocities. This results of this $V_D^{PF}$ fitting are shown in Fig.~\ref{fig:k_V_fitting} for Al-Cu and Si-As alloys. The relevant material properties for both alloys are given in Table \ref{table:materialProperties}.  
Here, the chosen fitting process yields reasonable agreement with the two experimentally calibrated $k^{CGM}(V)$ curves over the considered range of velocities. 
It is noted that the asymptotic analysis is formally most valid at small interface speeds, and thus excellent agreement could be achieved if \E{eq:k_V_PF} is matched to the the Aziz formula in \E{eq:k_V_CGM} only over small speeds, for example over $0<V<1$ m/s, which is still large enough to cover most rapid solidification experiments. 
%%%%
\begin{table}
\caption{Material properties for Al-0.5at\%Cu and Si-9at\%As. \\ *: $V_D^{PF}$ determined in Fig.~\ref{fig:k_V_fitting} }
\begin{tabular}{lcc}
                                                 & \multicolumn{1}{l}{Al-Cu} & \multicolumn{1}{l}{Si-As} \\ \hline
Equil. partition coeff. $k_e$  & 0.15\cite{Smith1994}  & 0.3 \cite{danilov2006}   \\
Melting point {[}K{]}                            & 933.3    &1685\cite{danilov2006}   
\\
Equil. liquidus slope $m_l^e$ {[}K/at\%{]}  & -5.3     & -4.0 \cite{danilov2006} 
\\
Alloy concentration $c_l^o${[}at\%{]}    & 0.5      & 9  
\\
Gibbs-Thomson coeff. $\Gamma$ {[}K m{]}     & 2.41e-7  & 3.4e-7 
\\
Liquid diff. coeff. $D_L$ {[}1e-9 m$^2$/s{]} & 4.4  \cite{Smith1994} & 15 \cite{Kittl2000}  
\\
Solid diff. coeff. $D_s$ {[}m$^2$/s{]} & 0                           & 0                       \\
Kinetic coeff. $\beta_0$ {[}s/m{]}          & 1.0 \cite{nath2017molecular} & 0.595\cite{Kittl2000} 
\\
Capillary anisotropy strength $\epsilon_c$       & -   & 0.03
\\
Kinetic anisotropy strength $\epsilon_k$       & -  & 0  
\\
Diff. velocity $V_D^{CGM}$ in \E{eq:k_V_CGM} {[}m/s{]}      & 6.7 \cite{Kittl2000}                             & 0.68 \cite{danilov2006}                         
\\
Diff. velocity $V_D^{PF}$ in \E{eq:k_V_PF} {[}m/s{]}$^*$      & 2.00                             & 0.385                          
\end{tabular}
\label{table:materialProperties}
\end{table}
\subsection{Phase field model convergence to continuous growth model} 
This section will show that the phase field model converges to the imposed partition coefficient $k^{PF}(V)$ given by \E{eq:k_V_PF}, and the CGM liquid-side concentration given by \E{eq:c_L_CGM}. It should be noted that the paper itself only extracts results required to map the phase field model equations onto the SIM described by CGM; the reader is referred to the supplementary material for detailed derivation of the matched asymptotic analysis from which these results were extracted.

All simulations were conducted with explicit Euler forward time stepping, with the time step size set to $0.7$ of the linear stability limit for the concentration diffusion equation. The phase field evolution in \E{eq:PF_phi_evolution} was solved with finite difference method, and the concentration diffusion equation in \E{eq:PF_c_evolution} with finite volume method. The mesh was adaptively refined to capture gradients in phase field and concentration fields appropriately with the software platform introduced in \cite{Greenwood18}, with the smallest allowed grid spacing set to 60\% of the interface width, $dx = 0.6 W_0$. 
%{\color{red}(Tatu: was this done with Mike's code or the 2D code, add a reference)}. 
The 1D runs assumed a constant dimensionless undercooling $\Delta = (T_l - T)( \, (1-k_e) |m_l^e| c_l^o \, )$. Capillary length magnitude was calculated as $d_o^{mag} = \Gamma / (\, (1-k_e) |m_l^e| c_l^o \,)$, using material properties from Table \ref{table:materialProperties}.

We studied the phase field model convergence to the corresponding CGM sharp interface model by measuring the instantaneous interface velocity, together with solid- and liquid-side concentrations during a 1D solidification following quenches to a fixed undercooling. 
The phase field model results reported below are shown to converge to an imposed $k(V)$ curve and CGM interface kinetic undercooling under transient conditions. 
%This is done to better approximate the situation prevalent in most rapid solidification experiments.
The transient conditions are considered so as to better approximate the situation prevalent in most rapid solidification experiments.
As a consistency check, the results reported here have also been validated under the more traditional steady-state conditions with a fixed thermal gradient, done using one-dimensional flat interfaces. 

Phase field runs without solute trapping ($k(V) = k_e$), for the classic sharp interface model (SIM) limit, were done with the kinetic coefficient $\beta$ fixed to either zero or to the literature-given value in Table \ref{table:materialProperties}, by setting $\tau$ based on \E{eq:beta_thinInterfaceRelation}. Phase field run with solute trapping and CGM kinetics were done with $\beta$ fixed to the literature-given value using \E{first_beta2}, where full solute drag used $a_2^-$ and zero solute drag used $a_2^+$ in the liquid-side concentration of $c_L^{CGM}$; in these cases, the partition coefficient $k(V)$ was set to follow $k^{PF}(V)$ in \E{eq:k_V_PF}. In total, we extracted data from phase field simulations with non-equilibrium conditions corresponding to four different cases:
%%%%%
\begin{align}
\text{Case 1}& \text{ (star): } 
k(V) = k_e, \text{$c_L$ from \E{eq:c_L_CGM} with $\beta = 0$ ,}
\nline
&\text{ and $f(k(V)) = 1$   }
\nline
\text{Case 2}& \text{ (square): }
k(V) = k_e  \text{, $c_L$ from \E{eq:c_L_CGM} with $\beta > 0$,}
\nline
& \text{ and $f(k(V)) = 1$  } 
\nline
\text{Case 3}& \text{ (triangle): }
k(V)  \text{ from \E{eq:k_V_PF}, $c_L$ from \E{eq:c_L_CGM}, $\beta > 0$,}
\nline
&\text{ and $f(k(V))$ from \E{eq:f_k_V_function} with no drag ($\mathcal{D} = 0$) } 
\nline
\text{ Case 4} & \text{ (circle): }
k(V)  \text{ from \E{eq:k_V_PF}}, \text{$c_L$ from \E{eq:c_L_CGM}, $\beta > 0$,}
\nline
&\text{ and $f(k(V))$ from \E{eq:f_k_V_function} with full drag ($\mathcal{D}= 1$)}, \nonumber
\end{align}
where the marker type for each case is shown in brackets (star, square, triangle, circle).
%%%

\begin{figure*}
\includegraphics[width=1.0\textwidth]{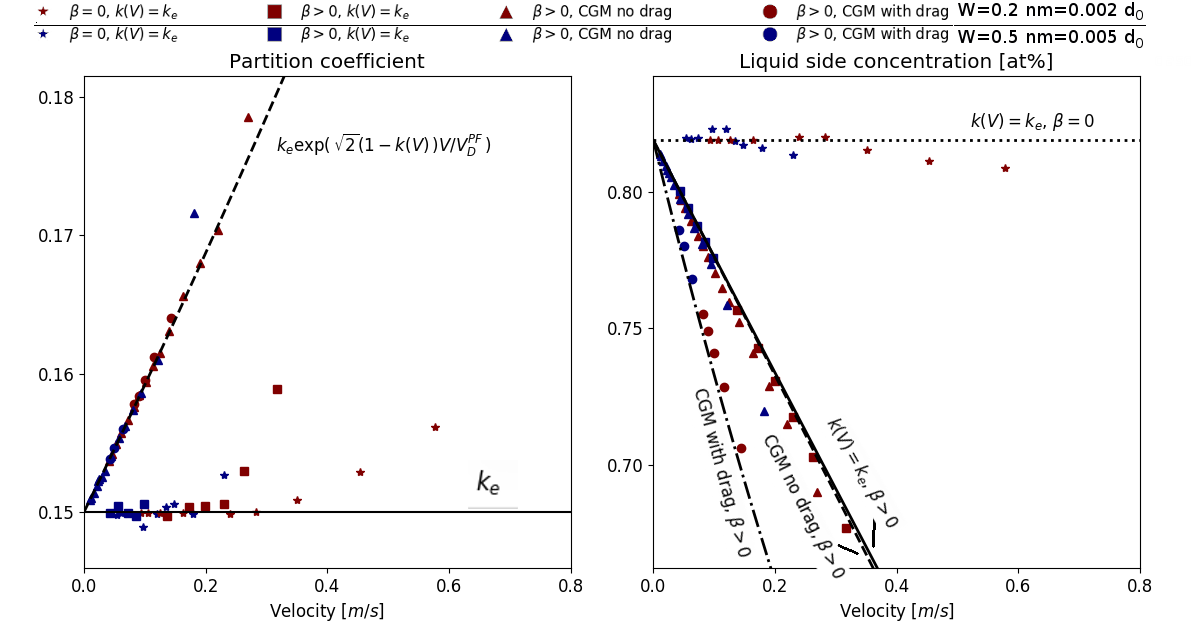}
\caption{Convergence of different phase field simulations  (red and blue scatter points) to corresponding  sharp interface models (solid and broken black lines) for an Al-Cu alloy, using material properties from Table \ref{table:materialProperties}.  The left graph shows convergence of the partition coefficient to $k_e$ and $k^{PF}(V)$ from \E{eq:k_V_PF}. The right graph shows the convergence of the liquid-side concentration $c_L^{CGM}$  to \E{eq:c_L_CGM} for the different non-equilibrium cases indicated in the text.  Each scatter point corresponds to an  instantaneous velocity and solid and liquid-side concentration measurement taken from the transient evolving of the concentration profile, using dimensionless undercooling $\Delta = 0.75$.}
\label{fig:AlCu_convergence}
\end{figure*}

%%%%FIG 3%%%
Figure~\ref{fig:AlCu_convergence} shows the convergence of the partition coefficient $k(V)$ (left graph) and liquid-side concentration $c_L$ (right graph) for the cases above, where material properties were taken for Al-Cu from Table \ref{table:materialProperties}. The data were obtained using two small computational interface widths $W$ to demonstrate that the phase field model converges well to the aforementioned sharp interface models at higher interface velocities. The dark red data corresponds to smaller interface width $W = 0.2$ nm,  and dark blue to $W = 0.5$ nm. As expected, the smaller interface width data (dark red) converge to the corresponding theory at a higher interface velocity than the larger interface width data (dark blue).   
% It s noteworthy that $c_L^{CGM}$ for the classic SIM model with $\beta>0$ is nearly identical to that case of no drag, while the solute partitioning is very different for the two cases. 
The dimensionless undercooling for these runs was set to $\Delta = 0.75$.
%%%
%%%
In all cases shown in the figure the interface velocity decreases monotonically over time, and both the partition coefficient and the liquid-side concentration converge to the corresponding sharp interface model (solid and broken black lines) at the measured instantaneous velocity. 

\begin{figure*}
\includegraphics[width=1.0\textwidth]{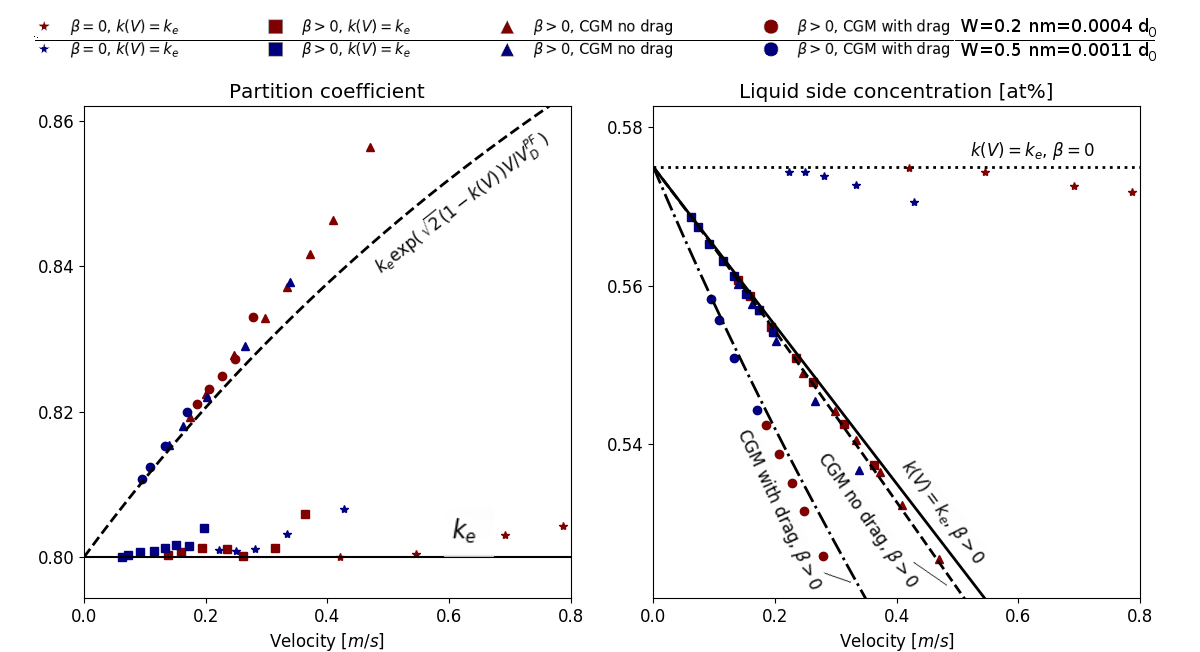}
\caption{Convergence of phase field models with different non-equilibrium features (red and blue scatter points) to the corresponding sharp interface models (solid and broken black lines) for Al-Cu alloy using material properties from Table \ref{table:materialProperties}, except $k_e$ is increased from $0.15$ to $0.8$. Left graph shows convergence to $k_e$ and $k^{PF}(V)$ from \E{eq:k_V_PF}. The right graph shows the convergence to liquid-side concentration $c_L^{CGM}$ to \E{eq:c_L_CGM} for the different CGM cases indicated. Dimensionless undercooling is set to $\Delta = 0.75$.}
\label{fig:AlCu_ke08_convergence}
\end{figure*}

%%%FIG 4%%%%
Fig.~\ref{fig:AlCu_ke08_convergence} compares phase field simulations with the same four non-equilibrium conditions as in Fig.~(\ref{fig:AlCu_convergence}) using the same Al-Cu material properties from Table \ref{table:materialProperties}, except that the equilibrium partition coefficient is increased from $k_e = 0.15$ to $k_e = 0.8$. For this increased partition coefficient, the convergence properties of the data are very similar to the original Al-Cu case shown in Fig~\ref{fig:AlCu_convergence}.
%%%%

\begin{figure*}
\includegraphics[width=1.0\textwidth]{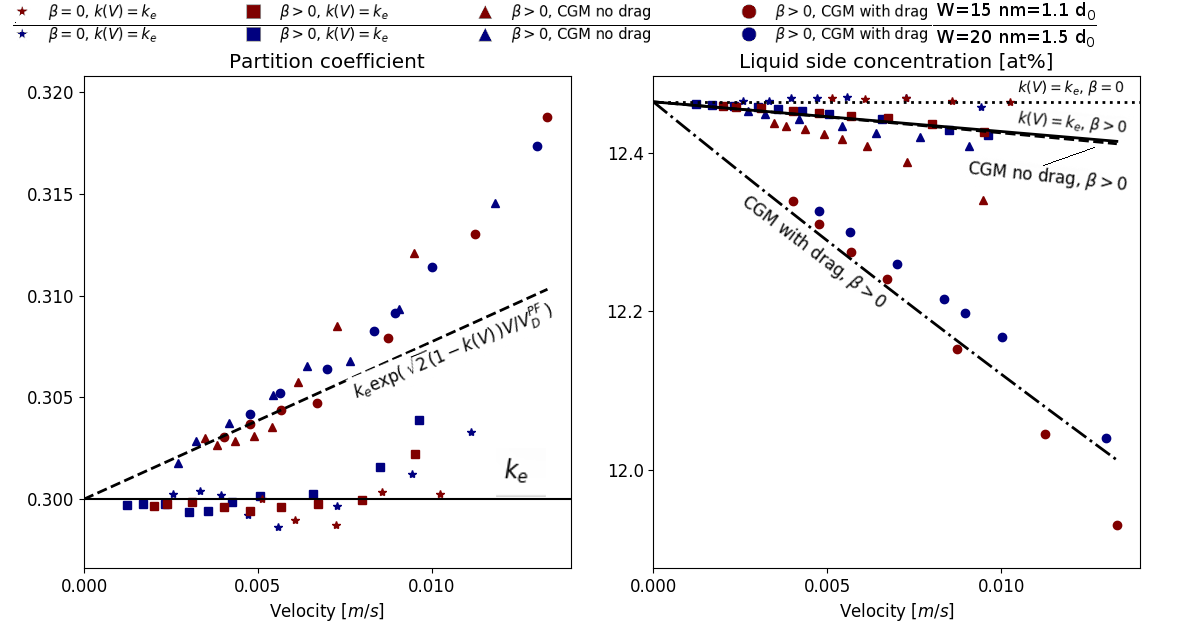}
\caption{ Convergence of different phase field simulations (red and blue scatter points) to  corresponding sharp interface models  (red and blue scatter points) for Si-Al alloy for two diffuse interface widths. The left graph shows convergence of the partition coefficient  to $k_e$ and  $k^{PF}(V)$ from \E{eq:k_V_PF}. The right graph shows the convergence of the liquid-side concentration $c_L^{CGM}$ to \E{eq:c_L_CGM} for the different non-equilibrium cases indicated in the text. Dimensionless undercooling is set to $\Delta = 0.55$.} 
\label{fig:SiAs_convergence}
\end{figure*}

%%%FIG 5%%%%%
Fig.~\ref{fig:SiAs_convergence} shows the same convergence as in Figs.~\ref{fig:AlCu_convergence} and \ref{fig:AlCu_ke08_convergence}, except material properties are taken for Si-As alloy in Table \ref{table:materialProperties}, and larger diffuse interface widths (scatter points in dark red correspond to  $W = 15$ nm, and in dark blue to $W = $ 20 nm); dimensionless undercooling is $\Delta = 0.55$. The phase field models converge excellently to the corresponding CGM (and classic) sharp interface models at low velocities. 
%{\color{red}(Tatu: was this a typo: "approximately 7 cm/s interface velocity."?)}. 
However, there is a larger relative scatter since the concentration projection error remains roughly the same as for the Al-Cu data in Figs.~\ref{fig:AlCu_convergence} and \ref{fig:AlCu_ke08_convergence}, but velocities are smaller, which implies that the partition coefficient $k(V)$ and liquid-side concentration $c_L^{CGM}(V)$ are closer to the equilibrium values at $V=0$. It is noted that phase field and sharp interface CGM models converge better at lower speeds since the asymptotic analysis is most accurate at low driving forces. however, this range of velocities is well within the scope of typical rapid solidification conditions.

It is noteworthy that the $c_L^{CGM}$ of the continuous growth model (CGM) without drag (black dash-dot line, case 3) is almost indistinguishable from the model using $k(V) = k_e$ and $\beta > 0$ (solid line, case 2) in Figs.~\ref{fig:AlCu_convergence}, \ref{fig:AlCu_ke08_convergence}, and \ref{fig:SiAs_convergence}. However, the difference between cases 2 and 3 become evident when comparing the solute partitioning $k(V)$.  

It should also be noted that estimating the solid-side concentration has the most scatter, when computed with the method described in section \ref{section:solidLiquidProjection}.  This is because for the transient concentration profiles, solid-side concentration has a complicated shape particularly in the initial stages of the simulation.
We also confirmed that the partition coefficient and $c_L^{CGM}$ from phase field simulations  converges to the corresponding sharp interface model when a thermal gradient and constant pulling speed is used; both under transient conditions, and when the interface reaches a steady state. These results are not shown in here to keep the length of the paper tractable. We chose to show results for constant undercooling and under transient conditions so as to better represent to experimental conditions where transient behavior can be important.

For the case of zero solute drag ($\mathcal{D} =0$), the use of extremely small interface width $W$ and diffusion velocity $V_D^{PF}$ (\E{eq:V_D_PF}) can make $a_2^+$ in \E{eq:a_2_nonEq} negative, which can eventually leads to a negative interface attachment time scale $\tau$ through sharp interface relation \E{first_beta2}, thereby making the model unphysical. In our experience this can become an issue only at small interface width $W$ which are not desirable in practical calculations.  

In all of the convergence graphs in Figs.~\ref{fig:AlCu_convergence}, \ref{fig:AlCu_ke08_convergence}, and \ref{fig:SiAs_convergence}, the significance of including non-equilibrium effects can be seen clearly. For Si-As alloy in Fig.~\ref{fig:SiAs_convergence}, already at 0.5 cm/s there is roughly a 5 \% relative difference between the concentration levels in equilibrium or non-equilibrium models of different cases; these differences can become magnified non-linearly in more complicated solidification conditions, such as two dimensional directional solidification presented in the following section.

\subsection{Demonstrating the effect of solute trapping on solidification  microstructure morphology}
\begin{figure}
\includegraphics[width=0.49\textwidth]{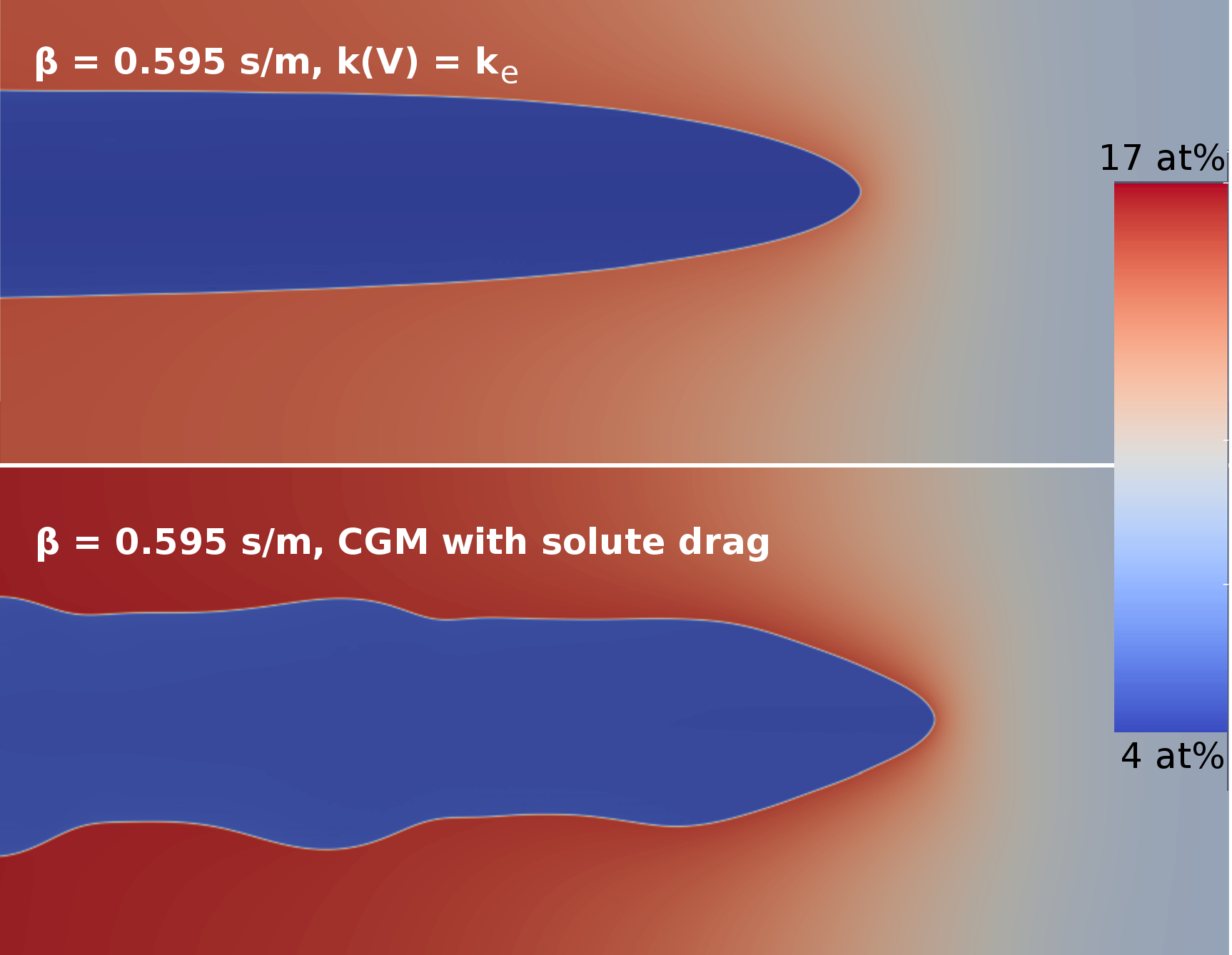}
\caption{Steady state directional growth with equilibrium partitioning (upper contour) versus solute trapping with solute drag (lower contour). Material parameters for Si-9at\%As alloy from Table \ref{table:materialProperties} in a thermal gradient of 400 000 K/m and pulling speed 0.5 cm/s. System size 12 $\mu m$ x 46 $\mu m$ in a co-moving reference frame. }
\label{fig:wide channel}
\end{figure}

This section demonstrates the significance of solute trapping in 2D solidification microstructure morphology. Directional growth of an Si-As alloy was simulated with parameters from Table \ref{table:materialProperties}. Steady state patterns of cellular growth fingers are shown in Fig.~\ref{fig:wide channel}. The figures plot a snapshot in time of the concentration  field. The upper contour corresponds to the case of no solute trapping (i.e., following the SIM with $k(V) = k_e$), while the the lower contour corresponds to the case of solute trapping with $k(V)$ from \E{eq:k_V_PF}, and with full solute drag ($\mathcal{D}=1$) according to \E{eq:c_L_CGM}. In both cases, the thermal gradient was set to 400 000 K/m, and the pulling speed was 0.5 cm/s. The simulations were done in a co-moving reference frame, with periodic boundary conditions in the vertical direction. The systems size was set to 12 $\mu m$ x 46 $\mu m$. We chose the pulling speed to be clearly smaller than the interface velocity where the transient 1D runs for Si-As converge in Fig.~\ref{fig:SiAs_convergence}. 

The data of Fig.~\ref{fig:wide channel} shows that for the two solidification conditions shown, including solute trapping (with full solute drag) leads to a thicker cell than for the case of no solute trapping. This can be motivated by the rough rule that microstructural length scale is inversely proportional to the material freezing range \cite{Trivedi94}; when solute trapping is active, the freezing range decreases. It is also noted that since the thicker cell in lower contour of Figure \ref{fig:wide channel} leaves less space to distribute the rejected solute in the liquid, this leads to a higher concentration levels in the remaining liquid.

The dendritic cell in the bottom simulation in Fig.~\ref{fig:wide channel} (which contains solute trapping) is seen to be growing side branches along the length of the trunk. This indicates that the cellular finger is becoming unstable, in contrast to the top frame of Fig.~\ref{fig:wide channel}, which remains cellular throughout the simulation. This instability is consistent with the fact that in directional solidification solute trapping can decrease the velocity where the growth mode changes from cellular to dendritic \cite{Trivedi94}.

%%%%%%%%%%
\section{Conclusion}
%%%%%%%%%%%
We presented a methodology, based on asymptotic analysis, for conducting quantitative phase field simulations of an alloy with controllable solute partitioning ($k(V)$) and a controllable  kinetic undercooling given by continuous growth model (CGM), tuned to follow either full or vanishing solute drag. The solute trapping model can be implemented to the standard ideal dilute binary alloy phase field model by applying two modifications: 1) in solute diffusion equation, replacing the standard antitrapping coefficient $a_t$ with a new coefficient that depends on an introduced \textit{trapping parameter}, and 2) in the sharp interface relation for kinetic coefficient $\beta$, replacing the  standard asymptotic analysis constant $a_2$ with a new constant that results into either complete solute drag or zero solute drag.

The phase field simulations were shown to converge to the intended sharp interface model in terms of the partition coefficient and the liquid-side concentration (which corresponds to a specific kinetic undercooling). The convergence was shown for various cases with different kinetic effects included: zero kinetic coefficient without solute trapping, non-zero kinetic coefficient without solute trapping, non-zero kinetic coefficient with solute trapping and without solute drag, and non-zero kinetic coefficient with solute trapping and with solute drag.

The phase field results were extracted by measuring the instantaneous interface velocity and solid- and liquid-side concentrations from a transient concentration profile under a fixed dimensionless undercooling. Similar results were found when extracting these measured quantities from a steady state moving interface pulled by a thermal gradient at a constant speed.

The considered phase field model was mapped onto the CGM limit with a matched asymptotic analysis for a general class of phase field models. 
This asymptotic analysis can be readily implemented to non-dilute and multicomponent alloys by using low supersaturation limit of a grand potential model, which can directly use the sharp interface relations as presented in this paper. 
%The modified phase field asymptotic analysis derived here can be readily implemented to non-dilute alloys by using low supersaturation limit of a grand potential model, which can be directly used in the sharp interface relations as presented in this paper. 

The presented phase field model with controllable solute trapping and CGM kinetics can be used to create more accurate process-microstructure maps for rapid solidification in order to, for example, determine morphological transition between dendritic, cellular, and planar growth in directional solidification. To properly model solute trapping and kinetic undercooling in simulations of industrially relevant applications, solute trapping measurements should be conducted for these respective alloys, for example for different grades of steels and nickel superalloys. 
\section*{Acknowledgement}
This work was supported by Academy of Finland under NANOSOLU project, VTT Technical Research Centre of Finland Ltd under the iBEX programme, National Science and Engineering Research Council of Canada, and the Canada Research Chairs. These funding sources are gratefully acknowledged.
\section*{References}
\bibliographystyle{elsarticle-num}
\bibliography{bibliomaster.bib}

\end{document}